# Student experiences in a computational physics lab through the lens of Physics Computational Literacy

Luke Nearhood  and Patti Hamerski
*Physics, Oregon State University, 301 Weniger Hall, Corvallis, OR, 97331*

Computational physics is a key part of what it means to do physics in the twenty-first century. However, upper division computational physics remains a largely understudied area. We set out to understand the experiences of students in an upper division computational physics lab course. To that end we conducted semi-structured interviews with five students at the end of their second of three terms in the course sequence. We then analyzed these interviews utilizing the emerging framework of Physics Computational Literacy. We found that the way students express how they learn computational physics implies they are making tradeoffs between their development of the different aspects of computational literacy. Additionally, we found that how students approach developing social computational literacy varies across individuals, and is driven by unspoken assumptions.

## I. INTRODUCTION

Computational physics is a key part of the landscape of twenty-first century physics [1, 2]. It is thus imperative that we understand how to effectively teach the subject so as to give physics students the tools they need to succeed in an evermore computerized world, without sacrificing the humanity necessary for building a strong physics community.

While research in computational physics education has expanded rapidly in recent years, there is still room for the development and application of theory in the context of computational physics education [3]. Further, upper division physics courses are still broadly understudied [4]. Therefore, upper division computational physics education is a particularly understudied and important area for research.

The place of computational physics in upper division undergraduate physics curricula has often been spotty[5]. This lack of comprehensive curricula coincides with a lack of research in this area.

## II. THEORETICAL FRAMEWORK

We initially approached this project as an open-ended endeavor. However, as noted above the need for theory in the area of upper division computational physics education prompted us to look to theoretical frameworks that could be elucidatory and that we could build upon.

Thus, the primary framework we have looked at our data through is that of Physics Computational Literacy (PCL) which was put forward by Odden *et al.* [6] based on diSessa's broader framework of computational literacy [7] and Berland [8]. Broadly, computational literacy is the notion that we can think of computing skills (coding, navigation, and using computers to solve problems) as a kind of literacy akin to reading, writing, or mathematical literacy.

In the framework of Physics Computational Literacy there are three main aspects to computational literacy, summarized in Table I.

Physics Computational Literacy is particularly well suited to the task of breaking the experience and process of learning computational physics into manageable pieces, yet still with deep theoretical implications. It also maintains the human centered element we are trying to keep in mind.

## III. CONTEXT & MOTIVATION

At Oregon State University we currently teach a sequence of three 10 week single credit lab courses at the junior level, which most students take alongside their other upper division physics courses in the paradigms program [9]. This course sequence consists of two 80 minute sessions per week. Students do the majority of coursework during class time.

The purpose of the course sequence is to teach students the basics of programming and how to apply those skills to physics. Students engage with the material through a series of Jupyter notebook assignments. Final group projects and individual midterm mini-projects were based on the structure of the computational essay [6].

| Computational Literacy Framework | |
|---|---|
| Literacy | Definition |
| Material (MCL) | How people practically use computing, such as through the act of coding. |
| Cognitive (CCL) | How people understand the computing they're doing, and how they use it to understand physics. |
| Social (SCL) | How people work together with computing, and communicate what they're doing and understanding. |

TABLE I. The three aspects of computational literacy, and our working definitions thereof.

The positionality of the authors is also an important piece of context. At the time this research was conducted, both authors were relatively new to the particular context of this university. The first author is a graduate researcher in physics, and not involved in teaching the course, giving him an outside perspective. This allowed him to develop rapport with students outside the constraints of a formal instructional environment. The interview environment was thus one of casual exploration and inquiry.

The second author is the instructor for the course, and as such is intimately familiar with the structures of the course and has more of an insider perspective, including an instructional relationship with students in the course. This role thus comes with some positional constraints, and as such this author was more involved in research design, providing perspective on the research context, and advising the overall coding process.

Our primary motivation at the outset was to understand student perspectives on and experiences with computational physics in the context of this course sequence. A secondary motivation was to characterize the effectiveness of a theoretical framework like PCL at explaining the learning that students do in this understudied setting. Our research questions for this paper were:

- What insights about student learning in upper division computational physics courses can we gain from looking at student experiences through the lens of Physics Computational Literacy?
- What insights does that give us about applying Physics Computational Literacy?

For the first question we are asking what students are saying about how they approach learning computational physics, as informed by the framework of PCL. For the second question, where does that potentially leave room for expanding on the theory.

## IV. METHODS

The first author conducted semi-structured interviews [10] with five students at the end of their second terms in the sequence (winter 2024). Conducting the interviews at this point in the course sequence provided students the opportunity to reflect on the prior terms, while also looking forward to the upcoming term, thus giving us a snapshot of their experience with the computational physics lab curriculum.

These interviews averaged 40 minutes in length. Examples of questions we asked include:
- If we were in class, what would your interactions with other students look like? Paint me a picture.
- If you were to give advice for how to succeed in this course to a future student what would it be?
- What is your background with coding/programming prior to this course?
- What do you see as the connection between learning to code and learning physics? If you see a connection?

These questions touched on all the main areas of Physics Computational Literacy, material, cognitive, and social.

We then coded the transcribed interviews broadly following Joseph Maxwell's [11] system of organizational, substantive, and theoretical coding. First breaking the interviews up into chunks according to question or topic, then coding en vivo for emergent themes, and finally coding according to what area of computational literacy those themes tended to fit under.

Consistent with our IRB protocol, the second author as the course instructor, did not have access to the raw interview data until after the course had concluded. The first author was the primary qualitative coder, and he would confer with the second author to discuss where he was uncertain about a specific code, for excerpts where it was challenging to articulate a justification for a particular code, and how the data fit into the emerging theoretical picture.

Through the process of coding the interview data for PCL we were able to identify where students expressed experiences that aligned with PCL. From that identification we were then able to identify tensions between the different aspects of how students were developing their Physics Computational Literacy.

## V. RESULTS

We found that students expressed tensions between the development of the different aspects of Physics Computational Literacy, and that social computational literacy development and practices are highly variable.

### A. Inter-Literacy Trade-offs

The way students express how they learn computational physics implies they are making tradeoffs between their development of the different aspects of computational literacy.

The following excerpt comes after asking about how the student, Betty, works with others in the classroom. She described how she often worked with a group that she also worked with in other physics courses.

> BETTY: I do sometimes feel like if I had... ***like a little bit more time to kind of think about what things are going on by myself that would like be good***... so I've decided to do like my final project for the class by myself to like, *think about the code a bit more*.

The emphasis Betty places on the notion of thinking about code and thinking about what is going on jumps out as a case of her emphasizing the importance of how to think about code. We can thus code this as an emphasis on knowledge of material computational literacy, what she knows about how to code.

Both of these factors, cognitive practices and material knowledge are then placed in contrast with working with her group. Group work, which is a practice we see as facilitating the development of social computational literacy, then in context the practice of developing her social computational literacy through group work is constraining her development of material and cognitive computational literacy.

Throughout the interview when discussing how her group works together Betty often emphasized a friendly competitiveness between the members.

> BETTY: Well my immediate goal with the assignment, the beginning of every class, I just want to beat my group members, ***I just want to be able to finish a part before they can***, so that I can tell them how to do it.

Here the emphasis on finishing parts of the assignment also speaks to an emphasis on material practices, as she wants to "get through" (a recurring phrase throughout the interview) the assignment. This would imply that for Betty in this context, the development of social computational literacy is facilitating the development of the skills necessary to "get through" the assignment, rather than cognitive computational literacy development.

We can also observe this at an earlier point in the interview.

> INTERVIEWER: Do you think that like the problem solving you have to do when coding relates to the problem solving you do in physics, or not?
>
> BETTY: I think it does, probably when you know how to code a little better. But at the moment since like getting the code to work is my general goal. *I feel like I think about the physics a little bit less and I think more about like how to get the thing to actually run a lot more.*

The distinction Betty is drawing here, between "getting the code to work" and "thinking about the physics" is an example of the distinction between material and cognitive computational literacy. In this case she is expressing a trade-off she has to make between focusing on getting the code to work and thinking about how she's using it to understand the physics. This connects back to the first excerpt where the resolution to this trade-off was to work on her own.

Betty was not the only student whose interview responses pointed toward these trade-offs.

Ethan also pointed towards a trade-off between group work and individual understanding when asked about group dynamics in the classroom. He discussed how he will usually work on his own at first.

> ETHAN: And I think I tend to like to work on things independently first, because *I think when you're able to think with yourself you think a little clearer*, it demands a little bit more, that you understand that you can't rely on other people to interpret it for you.

In describing how he waits to work with his partner or other neighbor when he gets stuck Ethan is drawing a connection between working alone and thinking about how to code clearly. Through the lens of computational literacy we can classify this as emphasizing cognitive computational literacy over social computational literacy, utilizing social computational literacy when it is conducive to accomplishing material tasks.

Alice, when asked what advice she would give to future students in the course, in the context of eliciting what norms and rules are present in the course, emphasized the importance of understanding over getting through all the assignment.

> ALICE: I would say that *understanding what you're doing is more important than doing it all*. ... Understanding how something works, understanding why it works is going to benefit you a lot more later in the class and in your future, um, than just finishing it all, getting the graphs and leaving. Because I used to do it all, like, finish the graphs and then leave.

Here Alice is explicitly stating that understanding is more important to her than just getting through the assignment. Drawing a distinction between "understanding why it works" and the product of getting the assignment done for which she uses the example of "getting the graphs" (the end product of many assignments). Through PCL we can classify this as prioritizing cognitive computational literacy in the form of understanding, over material computational literacy, in the form of finishing the assignments and getting the graphs.

### B. On The Social Nature of Computing

While in the prior section we discussed how student experiences represented trade-offs between the different aspects of computational literacy, we have also seen that how students approach developing social computational literacy varies across individuals, and is driven by unspoken assumptions.

Here Carl discusses how he has found more success this term when he's been working with someone who is at a similar level of coding knowledge to himself.

> CARL: So I would say that this term has been a little bit more comfortable because I found *someone that has about the same amount of coding knowledge that I do*. ... I mean, I learned the stuff, but I think this term, having someone *about the same approximate knowledge as to you*, working on stuff together it really helps.

Carl, in emphasizing the importance of having a similar level of knowledge about coding reveals how the development of social computational literacy can be facilitated by the way we structure groups. In this case by pairing students of similar levels of cognitive computational literacy and material computational literacy together.

Here, after being asked about group work in the class the student, Alice, shared how her experience with anxiety has influenced how she approaches group work in this context.

> ALICE: I have, an anxiety disorder. ... I'm good at coding. I know what I'm doing. I know how to ask for help. So I sit at a table alone. I talk to the TAs, we chit chat, I ask them questions. But I don't necessarily collaborate with my peers. ... *not because I don't believe that works. I believe that works for most people*. It just has never really worked for me. In the classroom, I see everybody collaborating. I see all the tables doing work. I see them talking about it. I see them working for the problems. And that's great. But that's not what I do. And it works great for me.

Additionally we see Alice allude to a tension between how she works, and her beliefs about what works for others, "I believe that works for most people." While having knowledge that her own practices work for her. That is a tension between what she seems to perceive about how social computational literacy ought to be developed and how she engages with it.

Alice, in discussing how her anxiety disorder has limited her participation in group work, provides us a window into how the development of social computational literacy can be underwritten by unspoken assumptions about what the social dimension constitutes. If we do not take into consideration the full range of activities encompassed under the umbrella of social computational literacy we may inadvertently limit student development thereof. This points toward the importance of accessibility and flexibility in how students develop social computational literacy.

The topic of unwritten rules and norms was also something we aimed to approach directly through the interviews. Here, Derek, discussed how the use of commenting and formatting code was not emphasized much in the term prior to when the interviews took place.

> INTERVIEWER: What would you say are like the norms of the classroom, or like unwritten rules?
>
> DEREK: I think not that it's like unwritten but we didn't really talk about it [commenting and formatting code] too much in 365. But the, like *etiquette* of like, code chunks, of putting in comments that wasn't super specified in 365, but it got brought up in [3]66.

The use of the term etiquette here also hearkens to a growing understanding of coding as an inherently social activity, as realized through commenting, formatting, and documentation.

## VI. DISCUSSION & CONCLUSIONS

From our analysis of the interview data we have presented here we have seen several key insights. First, Physics Computational Literacy [6] allows us to identify tensions and trade-offs between the different aspects of computational literacy. Second, students are conscious of those trade-offs that they may be making between the development of different aspects of computational literacy. And, social computational literacy development and practices can be bound up in unspoken assumptions.

Students are conscious of the trade-offs between learning key coding tools and interpreting how to combine them to solve physics problems. These two practices were coded for Material Computational Literacy and Cognitive Computational Literacy respectively, and the trade-offs articulated here signified where students have to negotiate a tension while developing the different computational literacies.

Social practices were shaped by often unspoken beliefs about the social nature of computing, such as Ethan's and Betty's beliefs that they can think about what they're doing more when they work alone. This implies to us that social computational literacy needs to be an explicit learning goal for computational physics courses.

However, as we saw with Alice, real time social interaction is not necessarily accessible to all students. And as such it is important to cultivate a learning environment that is conducive to collaboration, while also providing students some flexibility in how they engage in collaboration and thus the development of social computational literacy.

As we saw with Derek students can and do recognize the social nature of code when it comes to the importance of commenting and documenting code. And thus that real-time group work is not the only means of developing social computational literacy.

PCL is thus a useful framework for understanding teaching and learning computational physics. But it is also complicated and multifaceted. By focusing on where students are prioritizing the development of one aspect of PCL we can tailor future curricula more intentionally.

## VII. FUTURE WORK

We recently conducted more interviews with students in the computational physics lab curriculum which we will analyze further, in advance of redesigning the curriculum. These interviews were aimed at identifying areas where student learning expectations and experiences were disconnected from course learning outcomes.

Starting in the 2025-2026 academic year we will be updating our curriculum to consolidate the junior level computational physics courses into a single 3 credit course, and adding a course on physical computing and statistics at the sophomore level.

This redesign will provide us with the opportunity to incorporate lessons learned from this research and the literature into the curriculum.

One key lesson is the need for explicit instruction that targets the development of social computational literacy. This is in alignment with the literature on how social computational literacy has often been overlooked: "We see social computational literacy as an often neglected pillar within the framework That is, most computational exercises, projects, and curricula that we are familiar with give students frequent chances to program but do not often ask them to communicate with or about their code to people outside their working group" [6].

For example, this most recent winter term we piloted an assignment where students were asked to develop shared resources for overcoming common challenges they encountered in the computational physics course. This is an exercise primarily aimed at developing computational literacy skills which we hope to expand upon in the new version of the course.

Instruction also must balance structure and agency. For as we have demonstrated in the case of Alice, if we have too narrow a conception of how to develop social computational literacy, some students may be left out. However, if we do not explicitly design for the development of social computational literacy, then some students will be left without the support they need to develop it.

Further, we see a focus on tensions between the different aspects of Physics Computational Literacy as a useful way of applying and developing the theory. To that end we would encourage other researchers to try to apply this framework to other learning contexts to continue advancing the development of theoretical frameworks for computational physics education.